\documentstyle[preprint,aps]{revtex}

\draft

\begin{document}

\preprint{NCL94-TP15}

\title{Calogero-Vasiliev Oscillator in Dynamically Evolving Curved
Spacetime}

\author{J. W. Goodison \cite{email,www} }
\address{Physics Department, University of Newcastle upon Tyne,\\
Newcastle upon Tyne, NE1 7RU, U.K.}
\date{\today}

\maketitle

\begin{abstract}
  In a recent work, the consequences of quantizing a real scalar field
  $\Phi$
according to generalized ``quon'' statistics in a dynamically evolving
curved
spacetime (~which, prior to some initial time and subsequent to some
later time,
is flat~) were considered. Here a similar calculation is performed;
this time we
quantize $\Phi$ via the Calogero-Vasiliev oscillator algebra, described
by a
real parameter $\nu > -1/2$. It is found that both conservation ( $\nu
\rightarrow \nu$ ) and anticonservation ( $\nu \rightarrow - \nu$ ) of
statistics is allowed. We find that for mathematical consistency the
Bogoliubov
coefficients associated with the $i$'th field mode must satisfy $|
\alpha_i |^2
- | \beta_i |^2 = 1$ with $| \beta_i |^2$ taking an integer value.
\medskip
\begin{center}
To be published in {\bf Physics Letters B}
\end{center}
\medskip
\end{abstract}


\section{Introduction}

  Particle creation in an expanding universe is a well studied aspect
  of quantum
field theory in curved spacetime. Much of the original research in this
area was
performed by Parker~\cite{park1,park2,park3}; his papers concern the
interaction
of a quantized
field $\Phi$ with the classical background gravitational field arising
from a
spacetime dynamic between arbitrary times $t_1$ and $t_2$. As a
particular
example, we will choose the spatially flat ( $n$+1 ) dimensional
Robertson--Walker universe with metric

\begin{equation}
ds^2 = dt^2 - R^2(t) \delta_{ab} dx^a dx^b \quad \quad a,b = 1,2,
\dots, n.
\end{equation}
  with  constants $R_1$ and $R_2$
such that $R(t<t_1)=R_1$, ${R(t>t_2)=R_2}$. (~The spacetime portions
with
$t<t_1$ and $t> t_2$ are
usually referred to as the in- and out-regions respectively~). Parker
originally
demonstrated \cite{park2} that if $\Phi$ was a real scalar field of
spin zero
obeying the
Klein--Gordon equation; then if commutation relations were imposed upon
the
creation and annihilation operators in the in-region, it necessarily
followed that commutation relation must be imposed upon the creation
and
annihilation operators in the out-region. It is not consistent to
impose
anti-commutation relations upon such a field $\Phi$.

  Parker later generalized the proof to spin 1/2 fermions \cite{park3}
  which
obey the Dirac equation and satisfy anticommutation relations. The
cases of
higher spin bose/fermi fields, parastatistics \cite{park4} and ghost
fields
\cite{park5} have also been treated. Importantly, it was stressed that
the flat
space spin-statistics theorem could be inferred by taking the physical
limit in
which the dynamical part of the evolution becomes constant, and
demanding that
the commutation relations remain continuous in this limit.

  Recently, an attempt was made to generalize this work to ``quon''
  statistics
\cite{gootom}, a case which interpolates between conventional Bose and
Fermi
statistics. ( ``Quons'' are of great interest to theoretical physicists
and have
many suggested applications, for example
see~\cite{passal,wit,ger,kausal,iwa,solkat}~). It was discovered that
physically
consistent evolution of the ``quon'' algebra from the in-region to the
out-region took place only when the ``quon'' algebra reduced to the
special
cases of the Bose/Fermi algebras.

  It is now natural to ask if Parker's work may be applied to other
extensions of the Bose/Fermi algebra. Within this paper, we attempt to
quantize $\Phi$ via the Calogero--Vasiliev~{\cite{cal1,vas1} algebra

\begin{equation}
[ a, a^{\dagger} ] = 1 + 2 \nu K
\end{equation}
where $[A,B] = AB - BA$, $K$ is related to the number operator by $K =
(-1)^N$
and $\nu$ is a real parameter satisfying $\nu > -1/2$. We note further
that $N$
is given explicitly in terms of the
creation and annihilation operators by

\begin{equation}
N = { \{ a, a^{\dagger} \} \over 2 } - { ( 2 \nu + 1 ) \over 2 }
\end{equation}
  where $\{ A,B \} = AB + BA$; we shall also require the existence of a
  vacuum
state $|0>$ such that $a|0>=0$.

   Relation (2) is an obvious generalization of the Bose
algebra - it arises naturally in the
quantization of the simple harmonic oscillator experiencing an
additional
inverse square potential - and has been the subject of much interest
recently~(~for example, see
\cite{chajag,tur,brzegumac,brihanvas,yanzha,pol}~). In particular
Macfarlane
\cite{macca1,macca2} has
explored the links between the Calogero--Vasiliev system and
parastatistics of
order $p = 2\nu + 1$, potentially of great use since (2) is a bilinear
relation
as opposed to the trilinear relations usually associated with
parastatistics. It is these links that motivate us to study the
Calogero--Vasiliev oscillator in a dynamically evolving curved
spacetime.

  The structure of this paper then is as follows. In section II, we
  examine the
evolution of a field $\Phi$ which is decomposed in terms of in- and
out-region
creation and annihilation operators, where each individual mode
satisfies
relation (2). ( We will assume that different modes commute ). We find
that this
can be performed in a physically consistent way provided that the usual
Bose
condition $|\alpha_i|^2 - |\beta_i|^2=1$ must is satisfied for each
mode, where
$\alpha_i$ and $\beta_i$ are the ( diagonal ) Bogoliubov coefficients
linking
the in- and out-region operators. In section III, we obtain the
unexpected
result that $| \beta_i |^2$ must be an integer. Concluding remarks are
given in
section IV.

\section{Calogero--Vasiliev Oscillator in Dynamically Evolving Curved
Spacetime}

  As stated within the introduction, we will impose upon the in-region
  the
following commutation relations:

\begin{equation}
[ a_i, a_j^{\dagger} ] = \delta_{ij} ( 1 + 2 \nu K_i )
\end{equation}

\begin{equation}
[ a_i, a_j ] = 0
\end{equation}
  where $K_i = ( -1 )^{N_i^{in} }$ and $\nu$ is a real parameter which
defines the statistics of the in-region; different modes are assumed to
commute.
We shall
also assume that $\nu \not = 0$ since if that were the case, relations
(4) and
(5) would be the canonical commutation relations describing bosons -
Parker's
original work \cite{park2} would then force us to impose Bose
statistics upon
the out-region.

  We may now decompose the field operator $\Phi( x^{\mu} )$ ( where
  $\mu =
0,1,2, \dots n$, $x^0 = t$ ) in terms of the
in-region creation and annihilation operators obtaining

\begin{equation}
\Phi ( x^{\mu} ) = \sum_i [ F_i( x^{\mu} ) a_i + F_i^* ( x^{\mu} )
a_i^{\dagger}
]
\end{equation}
 where we regard $\Phi( x^{\mu} )$ as a real scalar field within the
 Heisenberg
picture ( since there will be nothing inherently relativistic about
this
calculation, a similar treatment could be given for the Schr\"odinger
field ).
$\{ F_i ( x^{\mu} ) \}$ is a complete set of positive frequency
solutions to the
Klein--Gordon equation.

  A similar expansion may also be performed in the out-region:

\begin{equation}
\Phi ( x^{\mu} ) = \sum_i [ G_i( x^{\mu} ) b_i + G_i^* ( x^{\mu} )
b_i^{\dagger}
]
\end{equation}
  where $ \{ G_i ( x^{\mu} ) \}$ is also a set of positive frequency
  solutions
to
the Klein--Gordon equation and the operators $b_i$ and $b_i^{\dagger}$
are in
general different from the $a_i$ and $a_i^{\dagger}$ due to particle
creation in
the expanding universe. We take the out-region algebra to be

\begin{equation}
[ b_i, b_j^{\dagger} ] = \delta_{ij} ( 1 + 2 \omega L_i )
\end{equation}

\begin{equation}
[b_i, b_j] = 0
\end{equation}
  with the real parameter $\omega$ but not necessarily equal to $\nu$,
and $L_i = (-1)^{N_i^{out} } $. We note that $\omega \not = 0$ since
otherwise
one could reverse the direction of time in Parker's original work to
show that
this would force $\nu = 0$. Since both sets $ \{ F_i ( x^{\mu} ) \}$
and $ \{
G_i
( x^{\mu} ) \} $ are assumed to be complete, we may expand one in terms
of the
other viz:

\begin{equation}
G_i ( x^{\mu} ) = \sum_j [ \alpha_{ij} F_j( x^{\mu} ) + \beta_{ij}
F_j^* (
x^{\mu}
) ]
\end{equation}
  where the complex numbers $\alpha_{ij}, \beta_{ij}$ are known as
  Bogoliubov
coefficients. However, for spacetimes whose metric is of the form (1),
the
Bogoliubov coefficients are known to be diagonal. Hence, we restrict
ourselves
to

\begin{equation}
\alpha_{ij} = \delta_{ij} \alpha_i \quad \quad \beta_{ij}= \delta_{ij}
\beta_i
\end{equation}

  Using relations (6), (7), (10) and (11) one obtains the relation
between the in- and out-region operators:

\begin{equation}
a_i = \alpha_i b_i + \beta_i^* b_i^{\dagger}
\end{equation}

  We now substitute (12) and its hermitian conjugate into relation (4).
  After
using the defining relations for the out-region statistics we obtain

\begin{equation}
( | \alpha_i |^2 - | \beta_i |^2 ) ( 1 + 2 \omega L_i ) = ( 1 + 2 \nu
K_i ).
\end{equation}

  If it were not for the operator terms involving $L_i$ and $K_i$, this
  would be
the usual Bose condition $|\alpha_i|^2 - |\beta_i|^2=1$. However, we
shall show
that (13) is in fact this condition in disguise. We take the scalar
product of
(13) with the out-region vacuum state $|0, out>$ which satisfies $b_i
|0, out >
=0$. We find that

\begin{equation}
( | \alpha_i |^2 - |\beta_i |^2 ) ( 1 + 2 \omega ) = 1 + 2 \nu
<0,out|K_i|0,out>
\end{equation}

  It becomes necessary at this point to calculate the value of
$<0,out|N_i^{in}|0, out>$. Using (3) and the transformation (12) we may
rewrite
the in-region Number operator as

\begin{equation}
N_i^{in} =  ( | \alpha_i |^2 + | \beta_i |^2 ) [ N_i^{out} + { ( 2
\omega + 1 )
\over 2 } ] + \alpha_i \beta_i b_i b_i + \alpha_i^* \beta_i^*
b_i^{\dagger}
b_i^{\dagger} -  { ( 2 \nu + 1 ) \over 2  }.
\end{equation}
Thus,

\begin{equation}
<0, out|N_i^{in}|0, out> = ( | \alpha_i |^2 + | \beta_i |^2 ){ ( 2
\omega + 1 )
\over 2 } - { ( 2 \nu + 1 ) \over 2 }.
\end{equation}

  Crucially, this is a real number. Therefore for equation (14) to
  remain
entirely real~(~since we assume $\nu \ne 0$ ), $<0,out| K_i |0, out>$
may only
take the values of plus or minus one. Hence, we have two alternatives:

\begin{mathletters}
\begin{equation}
<0,out|K_i|0,out> = +1 \quad \Rightarrow \quad | \alpha_i |^2 -
|\beta_i |^2 = {
1 + 2\nu \over 1 + 2 \omega }
\end{equation}

\begin{equation}
<0,out|K_i|0,out> = -1 \quad \Rightarrow \quad | \alpha_i |^2 -
|\beta_i |^2 = {
1 - 2\nu \over 1 + 2 \omega }
\end{equation}
\end{mathletters}

  We may also take the scalar product of (13) with the normalized
  $1$-particle
state $|1_i,out> \sim b_i^{\dagger}|0,out>$. We obtain

\begin{equation}
( | \alpha_i |^2 - |\beta_i |^2 ) ( 1 - 2 \omega ) = 1 + 2 \nu <1_i,
out|K_i|1_i,out>
\end{equation}
  where, again, to keep everything real, we must have that $<1_i,
out|K_i|1_i,out> =
\pm 1$. By adding (18) to (14) we discover that we have 3
possibilities:

\begin{mathletters}
\begin{equation}
<0,out|K_i|0,out> = <1_i, out|K_i|1_i,out> = +1 \quad \Rightarrow \quad
|
\alpha_i |^2 - |\beta_i |^2 = 1 + 2\nu
\end{equation}

\begin{equation}
<0,out|K_i|0,out> = <1_i, out|K_i|1_i,out> = -1 \quad \Rightarrow \quad
|
\alpha_i |^2 -
|\beta_i |^2 = 1 - 2\nu
\end{equation}

\begin{equation}
<0,out|K_i|0,out> = - <1_i, out|K_i|1_i,out>  \quad \Rightarrow \quad |
\alpha_i
|^2 - |\beta_i |^2 = 1
\end{equation}
\end{mathletters}

We may now rule out two of these possibilities. If we compare (17a)
with (19a),
we find that $\omega=0$. This contradicts one of our original
assumptions, and
so we discard this alternative. Similarly, comparing (17b) with (19b)
also gives
$\omega=0$.

  Our only remaining option then is to adopt $| \alpha_i |^2 - |\beta_i
  |^2 =
1$. Even then we have two choices; we may adopt $<0,out|K_i|0,out>=+1$;
then
from comparing (19c) with (17a) we conclude that $\nu = \omega$ i.e
conservation of statistics or we may adopt $<0,out|K_i|0,out>=-1$; then
comparing (19c) with (17b) we would conclude that $\nu = - \omega$
i.e.
anticonservation of statistics.

\section{Conservation and Anticonservation of Statistics}

  We now examine the situation more carefully for the case when $\nu
  =\omega$.
Consider the scalar product of the normalized $n$-particle states
$|n_i, out>
\sim ( b_i^{\dagger} )^n | 0, out >$.

\begin{eqnarray}
< n_i, out|N_i^{in}|n_i, out > &&= ( | \alpha_i |^2 + | \beta_i |^2 )[
n + { (
2 \nu + 1 ) \over 2 } ] - { ( 2 \nu + 1 ) \over 2 } \\
&&= n + | \beta_i |^2 ( 2n + 2\nu + 1 )
\end{eqnarray}
 where relation (21) follows from (15) and we have used the relation $|
 \alpha_i
|^2 - |\beta_i |^2 = 1$ in
obtaining
(21). We now use (21) in two special cases:

\begin{equation}
<0,out|K_i|0,out > = +1 \quad \Rightarrow \quad (-1)^{| \beta_i| ^2 ( 2
\nu + 1
)} =
+1
\end{equation}

\begin{equation}
<1_i, out|K_i|1_i,out > = -1 \quad \Rightarrow \quad (-1)^{ 1 + |
\beta_i| ^2 (
2 \nu + 3
) } = -1
\end{equation}

  Substituting (22) into (23) we obtain

\begin{equation}
(-1)^{ 2 | \beta_i |^2 } = +1
\end{equation}

  We deduce from this that $|\beta_i |^2$ must be an integer.  An
  almost
identical calculation reveals that $|\beta_i |^2$ must be an integer
when the statistics are anticonserved i.e. when $\nu = - \omega$.

  A direct consequence of this result is that in either case it is not
  possible
to take the physical limit in which the dynamical part of the evolution
goes to
zero; hence for this model the connection between spin and statistics
in flat
space may not be inferred.

\section{Discussion}

  We have shown that if a real scalar field $\Phi ( x^{\mu} )$ is
quantized according to relations (4) and
(5), then the particle statistics may either be conserved or
anticonserved upon
evolution from the in-region to the out-region ( depending on whether
$<0,out|K_i|0,out>$ is an even or odd integer respectively ). The usual
Bose
relation between
the Bogoliubov coefficients must hold for each field mode i.e. $|
\alpha_i |^2 -
|\beta_i |^2 = 1$. It is also necessary that $| \beta_i
|^2$ must be an integer for the evolution of the statistics to take
place in
a physically acceptable way.

  It is worth commenting upon the following form for the in-region
statistics ( a generalization to which relations (4) and (5) belong ):

\begin{equation}
[ a_i, a_j^{\dagger} ] = \delta_{ij} F( N_i^{in} ) \quad , \quad [a_i,
a_j ] = 0
\end{equation}
  where $F(r)$ is a function which returns a real number when the real
  variable
$r$ is an integer, and returns a complex number when $r$ is
non-integer. We
assume the number operator $N_i^{in}$ to be of the form

\begin{equation}
N_i = { \{ a_i, a_i^{\dagger} \} \over 2 } + const.
\end{equation}
  for a spacetime whose metric is of the form of relation (1). Similar
  relations
are supposed for the out-region:

\begin{equation}
[ b_i, b_j^{\dagger} ] = \delta_{ij} G( N_i^{out} ) \quad , \quad [b_i,
b_j ]= 0
\end{equation}
  where $G$ is a similar function to $F$ ( but not necessarily equal ),
  and we
wish to evolve the in-region
statistics to those of the out-region in a physically consistent way.
The
in-region and out-region creation/annihilation operators are still
linked by
(12). Taking the scalar product with the out-region vacuum state and
the
1-particle state $|1_i, out>$ yields

\begin{equation}
( | \alpha_i |^2 - | \beta_i |^2 ) G (0)  = F \bigl( < 0,out|N_i^{in}|
0,out >
\bigr)
\end{equation}

\begin{equation}
( | \alpha_i |^2 - | \beta_i |^2 ) G (1)  = F \bigl( <
1_i,out|N_i^{in}| 1_i,out
> \bigr)
\end{equation}
   In this case ( since $G(0)$ and $G(1)$ are real numbers ), to keep
   relations
(28) and (29) real both of the following must be satisfied:

\begin{equation}
< 0, out| N_i | 0,out > \in \{ 0,1,2,3,4,5,6,7,8......\}
\end{equation}

\begin{equation}
< 1_i, out| N_i | 1_i,out > \in \{ 0,1,2,3,4,5,6,7,8......\}
\end{equation}

  This in turn requires that $| \alpha_i |^2 + | \beta_i |^2 $ must be
  an
integer. Essentially this follows from chosen form of the number
operator.

\acknowledgements

  I wish to thank my supervisor David Toms for both suggesting this
  project and
for many helpful discussions along the way. I wish also to thank the
E.P.S.R.C.
for financial support.

\end{document}